\documentclass[12pt,a4paper,hypertex]{article}
\usepackage{jheppub}

\allowdisplaybreaks[1]
\usepackage{amsmath,bm}
\usepackage{subfigure}
\usepackage{color}

\begin{document}

\title{Self-consistent Analytic Solutions in Twisted $\mathbb{C}P^{N-1}$ Model 
in the Large-$N$ Limit}
\author[1,2]{Muneto Nitta} 
\author[3,2]{and Ryosuke Yoshii} 
\affiliation[1]{Department of Physics,  Keio University, 4-1-1 Hiyoshi, Kanagawa 223-8521, Japan}
\affiliation[2]{Research and Education Center for Natural Sciences, Keio University, 4-1-1 Hiyoshi, Kanagawa 223-8521, Japan}
\affiliation[3]{Department of Physics, Department of Physics, Chuo University, 1-13-27 Kasuga, Bunkyo-ku, Tokyo 112-8551, Japan}
\emailAdd{nitta@phys-h.keio.ac.jp}
\emailAdd{ryoshii@phys.keio.ac.jp}

\date{\today}
\abstract{
We construct self-consistent analytic solutions 
in the ${\mathbb C}P^{N-1}$ model in the large-$N$ limit, 
in which more than one Higgs scalar component take values 
inside a single or multiple soliton on an infinite space or on a ring, 
or around boundaries of a finite interval. }

\maketitle

\section{Introduction}

Nonlinear sigma models in two spacetime dimensions 
share a lot of non-perturbative properties with 
Yang-Mills theories or QCD in four dimensions,
such as asymptotic freedom, dynamical mass gap, 
dynamical chiral symmetry breaking
and instantons,
and therefore the former is regarded as a toy model of the latter 
\cite{Eichenherr:1978qa,Golo:1978dd,Cremmer:1978bh,Polyakov:1975rr,Polyakov:1975yp,Bardeen:1976zh,Brezin:1976qa,DAdda:1978vbw,DAdda:1978etr,Witten:1978bc}.
In particular, the ${\mathbb C}P^{N-1}$ model \cite{Eichenherr:1978qa,Golo:1978dd,Cremmer:1978bh} corresponding to the $SU(N)$ gauge theory 
has been studied extensively,  
such as the supersymmetric ${\mathbb C}P^{N-1}$ model 
\cite{Witten:1977xn, DiVecchia:1977nxl} 
for which the exact Gell-Mann-Low function was obtained \cite{Novikov:1984ac}
and dynamical mass gap was proved by the mirror symmetry \cite{Hori:2000kt}. 
The low-energy dynamics of a non-Abelian vortex string in a $U(N)$ gauge theory in four dimensions can be described by the ${\mathbb C}P^{N-1}$ model defined on a two-dimensional worldsheet
\cite{Hanany:2003hp,Auzzi:2003fs,Eto:2005yh} (see Refs.~\cite{Tong:2005un,Eto:2006pg,Shifman:2007ce,Tong:2008qd} as a review), 
giving a more precise correspondence between the ${\mathbb C}P^{N-1}$ model in two dimensions and the $U(N)$ gauge theory in four dimensions  \cite{Hanany:2004ea,Shifman:2004dr}.
\footnote{
The ${\mathbb C}P^2$ model appears also on a non-Abelian vortex  
in dense QCD \cite{Nakano:2007dr} (see Ref.~\cite{Eto:2013hoa} as a reivew).
}
The ${\mathbb C}P^{N-1}$ model on a compact direction with twisted boundary conditions has been also studied extensively.
In this case, an instanton is decomposed into several fractional instantons 
\cite{Eto:2004rz,Bruckmann:2007zh}.
Recently, bions which are composite states of fractional instanton and anti-instantons have been studied 
in the ${\mathbb C}P^{N-1}$ model 
for the application to 
the resurgence theory \cite{Dunne:2012ae,Misumi:2014jua}.
The ${\mathbb C}P^{N-1}$ model at finite density has been also studied \cite{Bruckmann:2016txt}.
Recently, there is a growing interest on 
the ${\mathbb C}P^{N-1}$ model defined on a finite region such as 
a ring \cite{Monin:2015xwa,Monin:2016vah}, 
a finite interval 
\cite{Milekhin:2012ca,Bolognesi:2016zjp,Betti:2017zcm,Flachi:2017xat}, 
and a disk \cite{Gorsky:2013rpa, Pikalov:2017lrb}.
In particular, the case of a finite interval may correspond to a non-Abelian vortex string 
stretched between two heavy non-Abelian monopoles 
\cite{Auzzi:2003em,Eto:2006dx}
or heavy non-Abelian monopole and anti-monopole \cite{Chatterjee:2014rqa}.

Since the ${\mathbb C}P^{N-1}$ model was defined forty years ago, 
only constant solutions have been studied for long time except for few cases:
the model on a finite interval studied recently 
around whose boundaries a Higgs scalar takes non-zero values 
\cite{Bolognesi:2016zjp,Betti:2017zcm,Flachi:2017xat}.  
Recently, in Ref.~\cite{Nitta:2017uog}, 
a class of self-consistent analytic solutions 
to gap equations has been obtained in the ${\mathbb C}P^{N-1}$ model 
in the large-$N$ limit in infinite space, 
by constructing a map from the gap equations in the ${\mathbb C}P^{N-1}$ model to those in the Gross-Neveu (GN) \cite{Gross:1974jv} or Nambu-Jona-Lasino 
\cite{Nambu:1961tp} model,
or equivalently to the Bogoliubov-de Gennes (BdG) equation.
The self-consistent analytic solutions that have been found 
include inhomogeneous Higgs configurations, 
such as a soliton in which a Higgs scalar field is localized, 
a lattice of such the soliton 
and multiple solitons with arbitrary separations, 
constructed from a real kink
\cite{Dashen:1975xh,Takayama:1980zz},
a real kink crystal 
\cite{Brazovskii,Yoshii:2011yt}, 
and multiple kink-anti-kink configurations 
\cite{Campbell:1981dc,Okuno:1983,Feinberg:2002nq,Feinberg:2003qz,Thies:2003br}, respectively in the GN model.
The appearance of the nonzero Higgs field, 
and consequently Nambu-Goldstone modes, 
are consistent with the Coleman-Mermin-Wagner theorem 
\cite{Coleman:1973ci,Mermin:1966fe} prohibiting
Nambu-Goldstone modes in two spacetime dimensions, 
because the Higgs field is confined in finite regions in our cases. 
The integrable structure \cite{Ablowitz:1974ry} behind the GN model 
\cite{Correa:2009xa,Takahashi:2012aw} and the map give infinite species of self-consistent analytic solutions.  
The self-consistent analytic solutions for the case of a finite interval have been also obtained \cite{Flachi:2017xat}  
from the corresponding solutions in the GN model \cite{Flachi:2017cdo}. 
The GN model on a ring was also studied numerically \cite{Yoshii:2014fwa} 
but the map to the ${\mathbb C}P^{N-1}$ model  was not applied to this case. 
However, all the previous works include only solutions in which only one Higgs field component takes a value 
and so are essentially solutions in the ${\mathbb C}P^1$ model.

In this paper, we construct self-consistent analytic solutions 
in which more than one Higgs scalar component take nonzero values. 
This is possible when we impose twisted boundary conditions or equivalently introduce a Wilson line for global flavor symmetry along spatial direction.
We first construct a single soliton and multiple solitons on an infinite space.
Around solitons, the Higgs phase appears with nonzero Higgs scalar components. 
We then construct a soliton on a ring. 
Finally, we construct the Higgs and confining phases in a finite interval.
For both cases, the Higgs field diverges around the boundaries.
For the former the Higgs fields are nonzero everywhere,
while for the latter the Higgs fields are nonzero almost everywhere except for the center of the interval. 
All of them are genuine solutions in the ${\mathbb C}P^{N-1}$ model with $N > 2$.
These solutions correspond to a non-Abelian vortex string stretched between 
non-Abelian monopole and anti-monopole whose orientational 
${\mathbb C}P^{N-1}$ modes are not aligned. 

This paper is organized as follows. 
In Sec.~\ref{sec:model} we define the $\mathbb{C}P^{N-1}$ model 
and give the gap equations in the large-$N$ limit and explain our method.
In Sec.~\ref{sec:solutions} we give several examples of 
self-consistent analytic solutions to the gap equation.
Sec.~\ref{sec:summary} is devoted to summary and discussion.

\section{Model and method}\label{sec:model}
In the following, we consider the $\mathbb{C}P^{N-1}$ model whose Lagrangian 
is given by
\begin{align}
\mathcal{L}&=
\int d^2 x \sum_{i= 1, 2, \cdots, N }\left[
D_\mu^0 n_i^\ast D^{0 \mu} n_i
-\lambda\left(|n_i|^2-r\right)\right]
\end{align}
where $n_i$ $(i=1,\cdots,N)$ are $N$ complex scalar fields, 
$D_{\mu}^0=\partial_\mu - i A^0_{\mu}$,  
and $A^0_\mu$ and $\lambda$ are auxiliary gauge and scalar fields, 
respectively. We set $A^0_\mu=0$ in the following. 
We impose a twisted boundary condition which generates a twist of the flavor degrees of freedom.
By picking up the first $m$ components, we consider a boundary condition under which $n_1,\cdots, n_m$ are twisted. 
The twisted boundary condition is equivalent to the presence of a Wilson line 
for a background $SU(m)$ non-dynamical gauge field. Then, the Lagrangian can be rewritten as
\begin{align}
\mathcal{L}&=
\int d^2 x \sum_{i\neq 1, 2, \cdots, m }\left[
\partial_\mu n_i^\ast \partial^\mu n_i
-\lambda\left(|n_i|^2-r\right)\right]\nonumber\\
&+\int d^2 x\left[(D_\mu \Sigma)^\dagger D_\mu \Sigma
-\lambda \Sigma^\dagger \Sigma\right],
\end{align}
where we have defined the $m$-component vector $\Sigma=(n_1, n_2 ,\cdots, n_m)^T$ and 
the covariant derivative $D_\mu=\partial_\mu-iA_\mu$ 
with the background non-dynamical gauge potential $A_\mu$ which is an $m\times m$ matrix and  $A_0=0$.  
The stationary condition for $\lambda$ and the $\Sigma^\dagger$ together with the 
Lorenz gauge $\partial_\mu A^\mu=0$ yields the gap equations 
\begin{align}
&[-\partial_x^2+\lambda(x)]f_n(x)=\omega^2_n f_n(x)\ (n\ge m), \label{eq1}\\
&\frac{N-m}{2}\sum_n\frac{|f_n|^2}{\omega_n}+\Sigma^\dagger(x)\Sigma(x)-r=0,\label{eq2}\\
&-D_x^2\Sigma(x)+\lambda(x)\Sigma(x)=0. \label{eq3}
\end{align} 
Here the $N-m$ factor appears in the second equation, whence the present calculation is valid in the sense of the $N-m$ expansion.
By using the redefinition of the field $\Sigma=\exp (i\int^x dx A)\tilde \Sigma $, one obtains 
\begin{align}
&[-\partial_x^2+\lambda]f_n=\omega^2_n f_n\ (n\ge m), \label{eq1-2}\\
&\frac{N-m}{2}\sum_n\frac{|f_n|^2}{\omega_n}+\tilde \Sigma^\dagger \tilde \Sigma-r=0, \label{eq2-2}\\
&-\partial_x^2\tilde \Sigma+\lambda \tilde \Sigma=0. \label{eq3-2}
\end{align}
In the following we choose the vector potential as $A=A^a_\mu T^a \otimes \mathbf{1}_{N-m, N-m}$, which induces the phase twist $i\int^L_0 A dx=\sum_i\gamma_i T_i$. 
Here $T_i$'s are $SU(m)$ generators.
We note that the ``gauge field " and the mass function $\lambda$ considered here 
becomes dynamical by considering the higher order corrections. 
However, we restrict ourselves to the leading order, in which these auxiliary fields do not become dynamical.

From now on, we focus on the case of $m=2$ for simplicity, 
though the following argument is straightforwardly applicable 
to arbitrary $m$. 
In the case of $m=2$, one can further rewrite the model as
\begin{align}
\mathcal{L}&=
\int d^2 x \sum_{i\neq 1, 2}\left[
\partial_\mu n_i^\ast  \partial^\mu n_i
-\lambda\left(|n_i|^2-r\right)\right]
+\int d^2 x\left(\partial_\mu \sigma^\ast \partial^\mu \sigma-\lambda  |\sigma|^2\right), 
\end{align}
where $\sigma=\tilde n_1+i\tilde n_2$ or $(\tilde n_1, \tilde n_2)=(\Re\sigma,  \Im \sigma)$ vice versa. 
Now the gap equations to solve become  
\begin{align}
&[-\partial_x^2+\lambda(x)]f_n(x)=\omega^2_n f_n(x)\ (n\ge 3), \label{eq1-3}\\
&\frac{N-2}{2}\sum_n\frac{|f_n|^2}{\omega_n}+|\sigma(x)|^2-r=0,\label{eq2-3}\\
&-\partial_x^2\sigma(x)+\lambda(x)\sigma(x)=0. \label{eq3-3}
\end{align}
In order to ensure the real eigenvalues for $\omega_n^2$, we consider $\lambda$ to be real 
such that $-\partial_x^2+\lambda$ to be Hermitian. 
Eq.\ (\ref{eq3-3}) describes a zero mode solution for the 
Schr\"odinger equation $(-\partial_x^2+\lambda)u=Eu$. 
By rewriting the $\lambda$ as 
\begin{equation}
\lambda=\Delta^2+\partial_x \Delta, 
\label{defDelta}
\end{equation}
one can solve Eq.\ (\ref{eq3-3}) as 
\begin{equation}
\sigma(x)\propto \exp \left[\int^x dy \Delta (y)\right].
\label{solsigma}
\end{equation}
We note that the other linearly independent solution is not normalizable. 
The remaining two equations (\ref{eq1-3}) and (\ref{eq2-3}) can be solved by the mapping given in the previous work (See Appendix A). 
The corresponding energy functional is given as 
\begin{equation}
E_{\mathrm{tot}}=(N-2)\sum_n\omega_n-r\int^\infty_{-\infty} dx (\Delta^2+\partial_x\Delta)
+ \left.\sigma\partial_x \sigma\right|_{-\infty}^{\infty}.\label{eq:energy}
\end{equation}

\section{Self-consistent analytic solutions on various spaces}\label{sec:solutions}
In this section we construct self-consistent analytic solutions on
an infinite space, a ring, and a finite interval in each subsection. 
\subsection{Infinite system}
We first consider the infinite size system. 
Some solutions for Eqs.~(\ref{eq1-3})--(\ref{eq3-3}) are given by \cite{Nitta:2017uog}
\begin{equation}
\sigma=0,\ \frac{A e^{i\phi}}{\cosh mx},\ \cdots \mathrm{etc}.
\end{equation}
Here we note that a localized Higgs soliton $Ae^{i\phi}\cosh^{-1} mx$ or a localized Higgs soliton lattice which we describe below are not inhibited from the Coleman-Mermin-Wagner theorem since these solutions do not have a long range order. 

The $\mathbb{C}P^1$ modes are given by 
\begin{equation}
\left(\begin{array}{c}
n_1\\
n_2
\end{array}
\right)=
e^{i\int^x A(y)dy}\left(\begin{array}{c}
\tilde n_1\\
\tilde n_2
\end{array}
\right)=
e^{i\int^x A(y)dy}\left(\begin{array}{c}
\Re \sigma\\
\Im \sigma
\end{array}
\right). 
\label{n12solgen}
\end{equation}
As an example,  if we choose $A=\alpha \sigma_x$, we obtain 
\begin{equation}
\left(\begin{array}{c}
n_1\\
n_2
\end{array}
\right)=
e^{i\alpha x \sigma_x}
\left(
\begin{array}{c}
\Re \sigma\\
\Im \sigma
\end{array}
\right)
=
\left(
\begin{array}{c}
\cos\alpha x\Re \sigma+i\sin \alpha x\Im \sigma
\\
\cos\alpha x\Im \sigma-i\sin \alpha x\Re \sigma
\end{array}
\right).
\end{equation}
For the case of $A=\beta \sigma_y$, we obtain 
\begin{equation}
\left(\begin{array}{c}
n_1\\
n_2
\end{array}
\right)=
e^{i\beta x \sigma_y}
\left(
\begin{array}{c}
\Re \sigma\\
\Im \sigma
\end{array}
\right)
=
\left(
\begin{array}{c}
\cos\beta x\Re \sigma+\sin \beta x\Im \sigma
\\
\cos\beta x\Im \sigma-\sin \beta x\Re \sigma
\end{array}
\right).
\end{equation}
Without loss of generality, we choose $A=\beta \sigma_y$ and $\Re \sigma=m,\ \Im\sigma=0$, yielding
\begin{equation}
\left(\begin{array}{c}
n_1\\
n_2
\end{array}
\right)=
m \left(
\begin{array}{c}
\cos\beta x
\\
-\sin \beta x
\end{array}
\right).
\end{equation}
In the case of $A=\gamma \sigma_z$, one obtains
\begin{equation}
\left(\begin{array}{c}
n_1\\
n_2
\end{array}
\right)=
e^{i\gamma x \sigma_z}
\left(
\begin{array}{c}
\Re \sigma\\
\Im \sigma
\end{array}
\right)
=
\left(
\begin{array}{c}
e^{i\gamma x}\Re \sigma
\\
e^{-i\gamma x}\Im \sigma
\end{array}
\right).
\end{equation}

\begin{figure}
\begin{center}
\includegraphics[width=20pc]{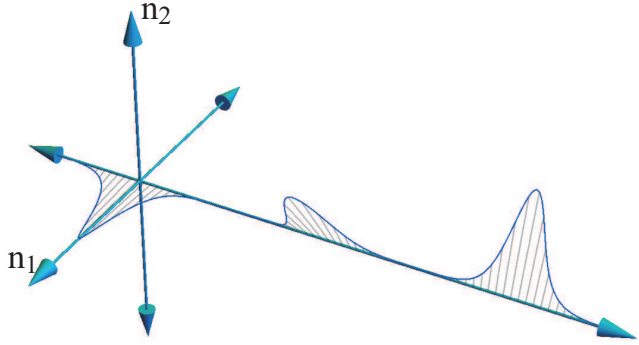}
\end{center}
\caption{The flavor winding localized Higgs lattice configuration with $A=\beta \sigma_y$. 
Here we set $\nu=0.99$, $\beta=3\pi/20\mathrm{K}(\nu)$, and $m=1$. 
The solution has bright soliton lattice configuration for the amplitude and the flavor rotates along the spatial axis.}
\label{Fig1}
\end{figure}

In Fig.\ \ref{Fig1}, we plot the Higgs lattice solution given by 
\begin{equation}
\left(\begin{array}{c}
n_1\\
n_2
\end{array}
\right)
\propto 
\left(-\sqrt{\nu}\mathrm{sn}(mx,\nu)+\mathrm{dn}(mx,\nu)\right)^{-1/\sqrt{\nu}}
\left(
\begin{array}{c}
\cos\beta x
\\
-\sin \beta x
\end{array}
\right).
\end{equation}
for $A=\beta \sigma_y$ which corresponds to $\Delta=\mathrm{sn}(x,\nu)$. 
Here $\mathrm{sn}$ and $\mathrm{dn}$ are the Jacobi's elliptic functions and $\nu$ is the elliptic parameter. 

\begin{figure}
\begin{center}
\includegraphics[width=35pc]{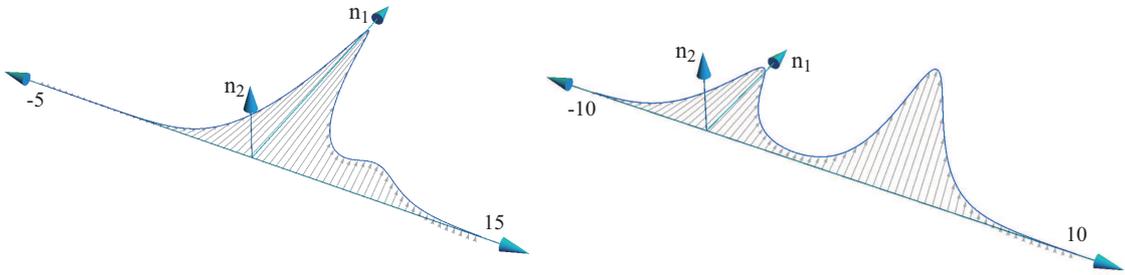}
\end{center}
\caption{The two soliton solution with different elliptic parameters with $A=\beta \sigma_y$. 
Here we set $\nu=0.9$ (left panel) and $\nu=0.999$ (right panel). We choose $\beta$ to be $\pi/30$.
We can observe that when we change the distance between the solitons, one rotates in the flavor space.}
\label{Fig1-2}
\end{figure}

In Fig.\ \ref{Fig1-2}, we plot a doubly localized Higgs soliton solution corresponding to 
\begin{align}
\Delta=k \tanh [kx-k\delta+R]
- \frac{ \omega_b e^{R}[\sinh (m_+ x-k\delta+2R)+\sinh(m_-x+k\delta)]}
{\cosh (m_+x-k\delta+2R)+ e^{2R}\cosh(m_-x+k\delta)}, 
\end{align}
where $\omega_b=\sqrt{m^2-k^2}$, $R=(1/2)\ln (m_+/m_-)$, and $m_{\pm}=m\pm k$. 
The left and right panels represent the twisted two soliton solution 
with $k=0.9$ (left) and $k=0.999$ (right), respectively.
The other parameters are set to be $\delta=0$ and $m=1$. 
By making the separation larger, the relative twist of the two peaks becomes larger. 

\subsection{Ring system} 
Next we consider the case of the $\mathbb{C}P^{N-1}$ model on a ring with the circumference $L$. 
In this case, the twisted boundary condition can be generated by the Aharonov-Bohm(AB)-like effect for the flavor degrees of freedom. 
We consider the periodic condition for all $\{n_i\}$. 
By the singular gauge transformation 
\begin{equation}
(\Sigma^T, n_3, \cdots, n_N)= (e^{i\int^x dy A} \tilde \Sigma^T, n_3, \cdots, n_N), 
\end{equation}
the twisted boundary condition 
\begin{equation}
\Sigma(x+L)=e^{i\beta} \Sigma(x)
\end{equation}
becomes  
\begin{equation}
e^{i\int^{x+L}_x dy A(y)} \tilde \Sigma(x+L)=e^{i\beta} \tilde \Sigma(x),
\end{equation}
whereas the background gauge field is completely eliminated from the self-consistent equations. 
Here $\beta$ is $2 \times 2$ Hermitian matrix. 
If the gauge field is chosen to compensates the twisting of the boundary, one can use the solutions obtained for the non twisting boundary conditions. 

In the case of the $SU(2)$ generator, we have  
$\exp (i\alpha \sigma_i x)=\cos \alpha x+i\sigma_i \sin\alpha x$. 
In the case of $A(y)=\alpha \sigma_i$, the above twisted boundary condition becomes 
\begin{equation}
\left(\cos \alpha L+i\sigma_i \sin\alpha L \right)\tilde \Sigma(x+L)=e^{i\beta}\tilde \Sigma(x).
\label{CodPhase}
\end{equation}
This shows the periodic structure on $\alpha$ which is similar to the AB oscillation effect. 
Because of this periodicity, $\alpha$ has $2\pi$ ambiguity. 
The lowest energy state corresponds to the case of $-\pi\le \alpha <\pi$ and the solutions with $\pi \le |\alpha|$ 
corresponds to the higher energy states. 
This can be easily shown as follows. 
If we move to the non-twisted boundary problem with the gauge potential, 
all of those solutions corresponds to the homogeneous solution. 
Thus the energy difference comes only from the gauge field part given by $\propto A^2$ and one can show the solution $(a)$ has the lowest energy. 

Apart from the phase winding induced by the boundary twisting, 
we have another condition for $\tilde \Sigma$ 
\begin{equation}
\tilde \Sigma(x+L)^\dagger \tilde \Sigma(x+L)=\tilde \Sigma(x)^\dagger \tilde \Sigma(x),
\label{CondAmp}
\end{equation}
since $e^{i\alpha\sigma_i L}$ is an unitary matrix. 
This means that the amplitude of the Higgs field also need to be periodic.

\begin{figure}
\begin{center}
\includegraphics[width=35pc]{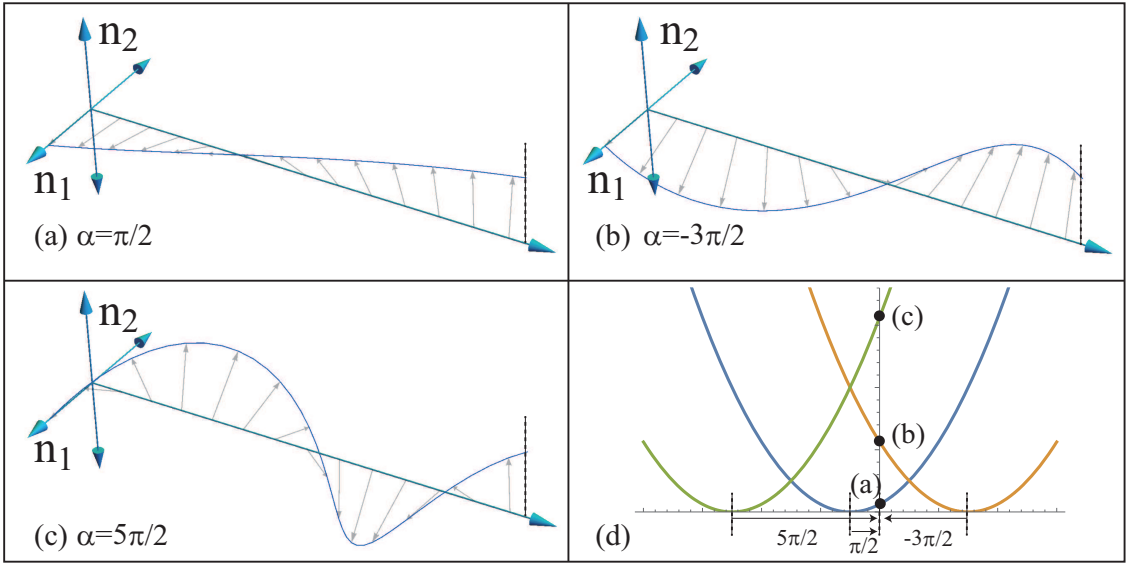}
\end{center}
\caption{The solution for the twisted boundary condition on the ring. 
Here we set $\alpha L=\pi/2$ (Fig.\ a), $-3\pi/2$ (Fig.\ b), and $5\pi/2$ (Fig.\ c). 
All of those solutions satisfies the same twisted boundary condition. 
In Fig.\ $(d)$, we depict the schematics of the phase winding. }
\label{Fig2}
\end{figure} 
In Fig.\ \ref{Fig2} we plot a homogeneous solution with the flavor rotation which corresponds to 
$\sigma=me^{i\phi}$. 
This solution is not allowed in the infinite system since the gap equation is not satisfied due to the 
absence of the infrared cutoff played by $1/L$ in the ring case. 
The absence of the long-range-ordered solution in the infinite system is consistent with the Coleman-Mermin-Wagner theorem. 
 
This solution obviously satisfies the condition (\ref{CondAmp}). 
The Fig.\ \ref{Fig2} $(a)$ shows the smallest winding solution.  
The higher winding states are also shown in $(b)$ and $(c)$ for the same boundary condition. 
The physical meaning of those solutions can be understood from the schematic figure $(d)$.  
Because the $2\pi$ periodicity of $\alpha L$, we have infinite branches of the solutions which gives the same twisting of the boundary. 
For example, the solution $(b)$ is the solution with the second smallest phase winding in which the flavor rotates opposite way compared with the solution $(a)$. 
The solution $(b)$ can be interpreted as the case of $\alpha L=\pi/2-2\pi$ 
which means that the solution $(b)$ belongs to the neighbor branch to the branch for $(a)$. 
In the same way, the solution $(c)$ is understood as the case of $\alpha L=\pi/2+2\pi$. 

\subsection{Finite interval system}
Finally we consider the case of the $\mathbb{C}P^{N-1}$ model on a finite interval which is relevant to describe the $\mathbb{C}P^{N-1}$ modes on a vortex string connecting heavy monopole and (anti-)monopole in the $U(N)$ gauge theory. 
If we have the flavor twisting between the opposite edges of the string, the solution (\ref{n12solgen}), 
which matches the twisted boundary conditions, becomes a solution. 
The other important point for this boundary condition is that the Higgs field $\Sigma$ unavoidably diverges at the edges. 

\begin{figure}
\begin{center}
\includegraphics[width=35pc]{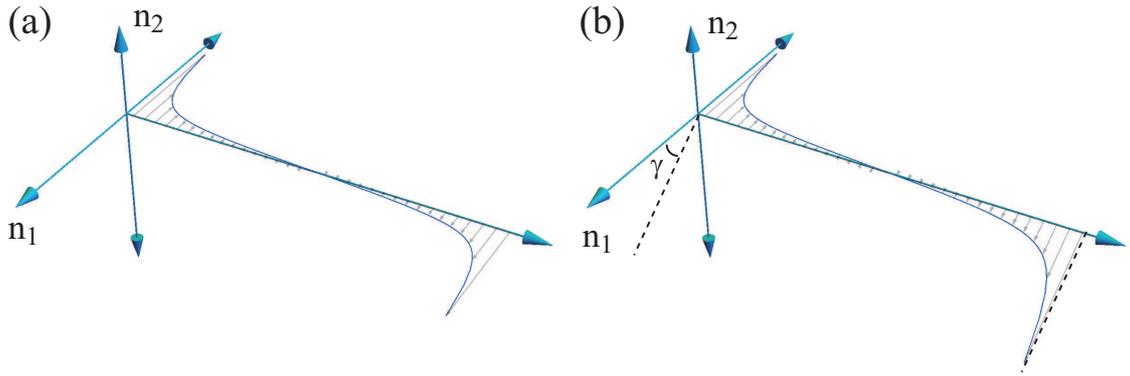}
\end{center}
\caption{The winding solution for the Dirichlet boundary condition corresponding to the confining phase. 
The left panel is an untwisted solution constructed in Ref.~\cite{Flachi:2017xat} while the right panel corresponds to the case with a twist of $\gamma=\pi/3$.
For the both cases we set $\nu=0$. 
The Higgs fields vanish at the center.
}
\label{Fig3}
\end{figure}

One of self-consistent analytical solutions of this problem can be given by 
\begin{equation}
\Sigma\propto \exp\left[\int dx \frac{4\mathrm{K}}{L\mathrm{sn}(4\mathrm{K}x/L,\nu)}\right] e^{i\gamma \sigma_i x/L} 
\left(
\begin{array}{c}
\cos \alpha \\
\sin \alpha
\end{array}
\right), 
\end{equation}
where we considered the twisted boundary condition on $\Sigma$ such as 
\begin{equation}
\Sigma(0)=\left(\cos \gamma+i\sigma_i \sin \gamma \right)\Sigma(L).
\end{equation} 
Here $\gamma$ is a constant parameter which characterizes the twisting of the boundary condition, $\nu$ is the elliptic parameter, and $\mathrm{K}(\nu)$ is the complete elliptic integral of the first kind. 
We also have another parameter $\alpha$ which could be set to $\alpha=0$ by a field redefinition since it corresponds to the overall $U(1)$ factor of the $n_1+in_2$. 
Thus, we obtain 
\begin{equation}
\Sigma\propto \exp \left[\int dx \frac{4\mathrm{K}}{L\mathrm{sn}(4\mathrm{K}x/L,\nu)}\right] e^{i\gamma \sigma_i x/L} 
\left(
\begin{array}{c}
1 \\
0
\end{array}
\right).
\end{equation}
This vanishes in the infinite size limit ($L\rightarrow\infty$, $\nu\rightarrow 1$). 
In the same limit, the mass gap function $\lambda$ becomes constant and thus this solution corresponds to the confining phase in the large size limit. 
We plot the confining phase solutions in Fig.~\ref{Fig3}.

\begin{figure}
\begin{center}
\includegraphics[width=35pc]{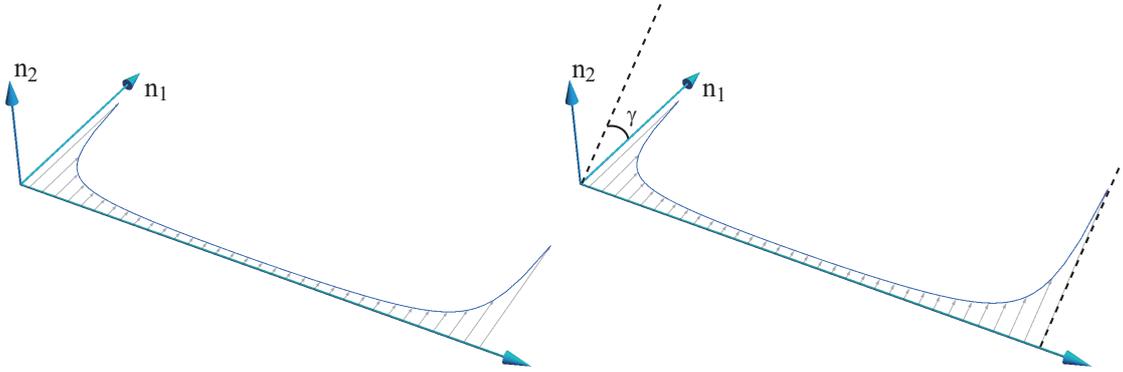}
\end{center}
\caption{The winding solution for the Dirichlet boundary condition corresponding to the Higgs phase. 
The left panel is an untwisted solution constructed in Ref.~\cite{Flachi:2017xat} 
while 
the right panel corresponds to the case with a twist of $\gamma=\pi/3$. 
For the both cases we set $\nu=0$. 
The Higgs field does not vanish anywhere in contrast to the case of the confining phase.
}
\label{Fig4}
\end{figure} 

There is another solution which corresponds to the Higgs phase in the infinite size limit 
where the Higgs field becomes the plane-wave like solution $\Sigma(x)=\text{const}\cdot \exp (i\sigma_i \tilde \gamma x) \cdot (1, 0)^T$: 
\begin{equation}
\Sigma\propto 
\exp \left[-\int dx \frac{2\mathrm{K}\mathrm{cn}(2\mathrm{K}x/L+\mathrm{K},\nu)}{L\mathrm{sn}(2\mathrm{K}x/L+\mathrm{K},\nu)}\right] 
e^{i\gamma \sigma_i x/L} 
\left(
\begin{array}{c}
1 \\
0
\end{array}
\right).
\end{equation}
We plot the Higgs phase solutions in Fig.~\ref{Fig4}.
This solution is inhibited in the infinite system since the gap equation is no longer satisfied in the limit. In other words, this solution is possible only in a finite system, to be
consistent with the Coleman-Mermin-Wagner theorem. 
For both solutions, the energy is scaled as $1/L$ \cite{Flachi:2017xat}.

\section{Summary and discussion}\label{sec:summary}
In the paper, we have constructed self-consistent analytic solutions of the ${\mathbb C}P^{N-1}$ model in the large-$N$ limit 
with the twisted boundary condition or 
equivalently with the background $SU(m)$ $(m<N)$ gauge field in the flavor space. 
The resulting solutions describe the various Higgs configurations with the $SU(m)$ flavor rotation.
 
In the present analysis, we have assumed that the flavor rotation is uniform since a nonuniform rotation costs more energy. 
However, nonuniform backgrounds could appear in a specific setup, which remains as a future problem. 

In this paper, we have used the mapping from the Gross-Neveu model to the $\mathbb{C}P^{N-1}$ model. 
It might be possible to generalize this mapping to the case of the chiral Gross-Neveu model, 
where the complex kink solution\cite{Shei:1976mn}, complex kink crystal solution \cite{Basar:2008im}, 
and the complex kink with arbitrary separation \cite{Takahashi:2012pk} have been obtained. 
In Ref.\ \cite{Nitta:2018yen}, a confining soliton in the Higgs phase was obtained, in which a confinement phase is localized
in the soliton core. This solution can be twisted as well. 

Our solutions could also be applicable to the condensed matter physics, for example, to the magnetic order 
\cite{Haldane:1982rj,Affleck,Senthil,Nogueira:2013oza}. 

In the case of a single component at a finite interval with the Dirichlet boundary condition, 
the Casimir force depending on the size of the system was discussed before 
\cite{Betti:2017zcm,Flachi:2017xat}, 
where the Casimir force gives either attractive or repulsive pressure to the system size.
For the case of the twisted boundary conditions studied in this paper, one can further discuss a Casimir force acting on the flavor internal space of ${\mathbb C}P^{N-1}$. 
In this case, 
the force gives either attraction or repulsion between the ${\mathbb C}P^{N-1}$ modes on the boundaries.
In the context of a non-Abelian string stretched between a monopole and (anti-)monopole, this force attains ferromagnetic or anti-ferromagnetic properties, respectively, on the monopole and (anti-)monopole.

The twisted boundary condition in the temporal direction has also been investigated. 
The large-N volume independence and the absence of the Affleck transition have been shown in the setup \cite{Sulejmanpasic:2016llc}. 
Our formalism and inhomogeneous solutions may possibly be used also in this case. 

During completion of this paper, we were informed that the authors of Ref.~\cite{Betti:2017zcm} were writing a paper on similar configurations. 

\section*{Acknowledgement}
We thank Sven Bjarke Gudnason, Kenichi Konishi and Keisuke Ohashi 
for correspondence.  
The support of the Ministry of Education,
Culture, Sports, Science (MEXT)-Supported Program for the Strategic Research Foundation at Private Universities `Topological Science' (Grant No.\ S1511006) is gratefully acknowledged. 
The work of M.~N.~is 
supported in part by the Japan Society for the Promotion of Science
(JSPS) Grant-in-Aid for Scientific Research (KAKENHI Grant
No.~16H03984 and 18H01217) 
and by a Grant-in-Aid for Scientific Research on Innovative Areas ``Topological Materials
Science'' (KAKENHI Grant No.~15H05855) from the MEXT of Japan. 

\appendix

\section{Mapping between the Gross-Neveu model and $\mathbb{C}P^{N-1}$ model}
In this appendix, we review the mapping between the GN model and the $\mathbb{C}P^{N-1}$ model \cite{Nitta:2017uog}. 
We consider the gap equations (\ref{eq1-3})--(\ref{eq3-3}) which have to be solved self-consistently.  
By using $\Delta$ defined in Eq.\ (\ref{defDelta}), 
the Klein-Gordon-like equation (\ref{eq1-3}) can be rewritten as 
the following Dirac-like equation 
\begin{align}
\left(
\begin{array}{cc}
0 & \partial_x +\Delta \\
-\partial_x+\Delta & 0
\end{array}
\right)
\left(
\begin{array}{c}
f_n \\
g_n
\end{array}
\right)
=
\omega_n
\left(
\begin{array}{c}
f_n \\
g_n
\end{array}
\right).
\label{eqBdG}
\end{align}
By eliminating $g_n$, one obtains Eq.\ (\ref{eq1-3}). 
The same procedure for Eq.\ (\ref{eq3-3}) yields 
\begin{align}
\left(
\begin{array}{cc}
0 & \partial_x +\Delta \\
-\partial_x+\Delta & 0
\end{array}
\right)
\left(
\begin{array}{c}
\sigma \\
\tau
\end{array}
\right)
=0.
\label{eqsigma}
\end{align}
This equation is nothing but an equation for zero modes. 
Thus, the solution is given by Eq.\ (\ref{solsigma}). 
Differentiating Eq.\ (\ref{eq2-3}) by $x$ and substituting the solution (\ref{solsigma}) into that, one obtains 
\begin{equation}
\Delta=\frac{N-2}{2r}\sum_n f_n g_n. 
\label{gapeqGN}
\end{equation}
This equation self-consistently determines $\Delta$ together with Eq.~(\ref{eqBdG}). 

Now the three gap equations reduces to the two equations (\ref{eqBdG}) and (\ref{gapeqGN}). 
The latter two equations coincide to the gap equations appearing in the GN model; 
the so-called Bogoliubov-de Gennes equation (\ref{eqBdG}) and the gap equation (\ref{gapeqGN}). 
By determining $\Delta$ from those equations, one can calculate $\sigma$ by Eq.~(\ref{solsigma}). 
It should be noted that the normalization of $\sigma$ must be fixed from Eq.\ (\ref{eq2-3}), 
since Eq.\ (\ref{gapeqGN}) is obtained from the differentiation of Eq.\ (\ref{eq2-3}) 
and thus the information for the normalization is lacking. 
For example, the constant solution $\Delta=m$ exists for Eqs.\ (\ref{eqBdG}) and (\ref{gapeqGN}) in the infinite system, 
but this solution cannot satisfy Eq.\ (\ref{eq2-3}).

\end{document}